# Broadband Mid-Wave Infrared Imaging Enabled by All-Silicon Hyperboloid Metalens

Rose Johnson[1,#] , Debabrata Mandal[2,#], Zihao Zou[2], Jessie Wei[3], Anna Wirth-Singh[3], Vladimir Yarmolik[1], Johannes Froech[2], Arka Majumdar[1,3,*] Praneeth Chakravarthula[2,*]

[1] Department of Electrical and Computer Engineering, University of Washington, Seattle, WA 98195, USA.

[2] Department of Computer Science, University of North Carolina at Chapel Hill, Chapel Hill, NC 27517, USA.

[3] Department of Physics, University of Washington, Seattle, WA 98195, USA.

[*] Corresponding authors. Email: arka@uw.edu, cpk@cs.unc.edu

[#] These authors contributed equally

**Abstract**

The mid-wave infrared (MWIR, 3 – 5 μm) is indispensable for thermal sensing, gas detection, long-range imaging, and free-space optical communications.[1,2] Meta-optics offer a compact, lightweight, and potentially low-cost alternative to conventional refractive optics, making them attractive for size, weight, and power (SWaP)-constrained systems. However, their strong chromatic dispersion has largely restricted imaging to narrow spectral bands.[3] Here, we demonstrate broadband imaging across the full 3–5 μm MWIR band using a single-layer silicon metasurface coupled with computational reconstruction. Remarkably, the reconstruction model is adapted using only four image captures, overcoming large training data requirements and enabling broadband image recovery despite the strongly wavelength-dependent optical response. In contrast to approaches that engineer complex, dispersion-optimized metalens profiles for broadband operation, our method leverages the wavelength-dependent information encoded by a simple 2-cm-diameter hyperboloid metalens. We show that this metasurface captures sufficient broadband intensity information for computational reconstruction across the entire MWIR band. These results establish a simple route toward broadband computational meta-optics without requiring broadband achromatic metalens designs, offering a promising platform for compact MWIR imaging systems.

## Introduction

Imaging in the mid-wave infrared (MWIR, wavelength band 3 – 5 μm) is central to a broad range of scientific, industrial, and defense applications, ranging from long-range imaging and remote sensing to free-space optical communication and adaptive optical wavefront sensing through atmospheric turbulence.[3–7] Conventional MWIR imaging systems typically rely on refractive lenses made from a limited set of infrared-transparent materials, such as germanium and chalcogenide glasses. These optics provide high image quality, but they are often bulky and heavy. This becomes a major limitation for platforms such as aircrafts, drones, and other systems where size, weight, and power (SWaP) are tightly constrained.

Flat meta-optics offer a fundamentally different approach. Instead of shaping light with a thick, curved lens, a metasurface uses a thin layer of subwavelength structures to control the optical wavefront.[8,9] This can dramatically reduce the volume and mass of an imaging system. MWIR metasurfaces have been demonstrated using several material platforms, including chalcogenide glasses such as ZnSe and ZnS, Si on sapphire or $CaF_2$, single-layer Si, and double-layer ZnS/Si structures.[2,3,10–12] Silicon is particularly attractive because all-silicon metasurface can be readily fabricated using high volume semiconductor manufacturing techniques.

However, the major obstacle to broadband meta-optic imaging is chromatic aberration. A conventional hyperboloid metalens is designed to focus one wavelength, and its focal response changes rapidly away from that design wavelength because of the strong wavelength dependence of its phase profile.[13,14] As a result, an image that is sharp at one wavelength can become severely blurred at another. This problem becomes especially challenging when a single metasurface must operate across a broad spectral band such as the full 3–5 μm MWIR window.

Two main approaches have emerged to address this challenge. The first is *dispersion engineering*, in which the individual meta-atoms are designed to produce the desired phase response across multiple wavelengths. In the visible, this approach has enabled broadband achromatic metalenses with diffraction-limited focusing, and related designs have been demonstrated at narrower bandwidths in the infrared.[15–21] However, the achievable bandwidth, numerical aperture (NA), and aperture size are fundamentally coupled.[13,22,23] As a result, many demonstrated achromatic metalenses operate with relatively small apertures or low NA, limited to only a few phase wraps. This limitation becomes increasingly important for practical imaging systems, where a moderate-to-large aperture can require thousands of phase wraps across the lens. In the MWIR, broadband metalenses have been experimentally demonstrated, but typically over only part of the full 3–5 μm band or at relatively small aperture and NA.[3,6,17]

The second approach is *computational imaging*, where rather than requiring the optics alone to form a sharp image at every wavelength, the optical design and reconstruction algorithm work together. For example, inverse-designed meta-optics jointly optimized with a computational backend using an image-quality metric such as the modulation transfer function (MTF) have been demonstrated over the desired spectral range.[24–26] Such designs generally trade the best possible performance at any single wavelength for a more uniform response across the full bandwidth. A computational backend then removes the remaining errors. This strategy has produced compelling broadband imaging results in the visible and has also been extended to the long-wave infrared (LWIR, 8 – 12 μm), where inverse designed all-Silicon meta-optics enabled broadband thermal

imaging.[24,27]

Interestingly, complex dispersion engineering is not always necessary for computational imaging. Broadband computational imaging has also been demonstrated in the visible using a conventional hyperboloid metalens, even though its point spread function (PSF) becomes highly distorted away from the design wavelength.[28] In that case, information shared among the red, green, and blue (RGB) channels was used by a learned reconstruction model to correct chromatic blur and vignetting. This raises an important question: *can a similarly simple metalens enable broadband computational imaging in the MWIR, where no RGB color channels are available?*

The answer is not obvious because visible and MWIR cameras sample spectral information differently. Visible cameras typically use RGB color filters, which divide the visible spectrum into narrower spectral channels.[29,30] Apart from preserving color information, these filters also reduce the spectral bandwidth that contributes to each measurement, effectively making the operational wavelength range narrower. In contrast, a broadband MWIR detector can collect photons across the entire 3–5 μm band in a single channel. The optics must therefore tolerate a much larger fractional bandwidth, making broadband optical optimization substantially more difficult. At the same time, MWIR intensity imaging does not necessarily require the system to preserve color information. Instead, the central requirement is to collect enough useful spatial information across the band for the scene to be reconstructed.

This distinction leads to a surprisingly simple design principle. We show that, across the large fractional bandwidth of the MWIR, an inverse-designed metalens can provide less useful imaging performance than a conventional hyperboloid metalens. Although the hyperboloid produces strong chromatic blur, it retains sufficient broadband spatial contrast and photon throughput for computational reconstruction. However, this poses a computational challenge wherein the raw measurements are blurred and must therefore be restored without relying on the thousands of paired training images commonly used to train neural networks.[27,27,31,32]

Here, we demonstrate broadband computational imaging across the full MWIR band using a 2 cm diameter, f/1 all-Silicon hyperboloid metalens designed for a nominal wavelength of 4.5 μm. We first compare the simulated broadband performance of the hyperboloid metalens with that of an inverse-designed metasurface at matched numerical aperture and across different spectral bandwidths. We find that as the bandwidth increases, the simple hyperboloid becomes the better optical front end for computational reconstruction. We then introduce an image restoration approach that conditions a pretrained restoration network on the experimentally measured MWIR PSF and adapts it to our imaging system using only four captured images (Fig. 1a). Our metalens itself uses simple meta-atoms and requires no dispersion engineering (Figs. 1b–d). Together, these results show that broadband MWIR imaging does not require a metalens that forms a sharp image at every wavelength. Instead, a simple chromatic metalens can capture sufficient spatial information for high-fidelity computational reconstruction across the entire MWIR band.

## Results

### Broadband information transfer favors a hyperboloid metalens

Unlike visible cameras, which typically use color filters to divide the spectrum into narrower RGB channels, MWIR cameras can collect light across the full 3–5 μm band in a single channel.

Therefore, we seek to investigate under what conditions a dispersion-engineered meta-optic preserve more spatial information than a simple hyperboloid metalens.

To this end, we calculated the integrated volume under the modulation transfer function (MTF) across varying fractional bandwidths ($\Delta\lambda/\lambda_0$) and report the results in Fig. 2. Because the MTF volume serves as a direct proxy for the spatial frequency information transferred to the sensor and available for computational restoration, this metric establishes the boundary conditions where engineering the chromatic dispersion yields a net advantage.

At an intermediate fractional bandwidth of $\sim 0.10$ - $0.20$, the inverse-designed meta-optic achieves a larger integrated MTF volume than the hyperboloid metalens. This regime is relevant to previous computational imaging demonstrations in the visible, where RGB Bayer filters divide the spectrum into channels with fractional bandwidths of $\sim 0.18$.[25,27] Within this bandwidth range, inverse design can balance the transfer of spatial frequencies across each spectral channel and improve the information available for reconstruction.

This behavior changes as the spectral bandwidth increases. Full-band MWIR imaging without spectral filters corresponds to a much larger fractional bandwidth of $0.50$. Over this broad spectral range, the inverse-designed meta-optic must balance its response across a larger range of wavelengths. As a result, its integrated MTF volume decreases below that of the hyperboloid metalens. At this operating point (fractional bandwidth $= 0.50$), the hyperboloid retains a 9% higher integrated MTF volume compared to the inverse-designed counterpart. Consequently, for full MWIR imaging, a standard hyperboloid metalens provides superior photon throughput and contrast transfer, making it the preferred optical front-end when paired with a computational backend.

**MWIR metalens design**

Guided by this analysis, we designed and fabricated a 2-cm-aperture hyperboloid metalens with a nominal focal length of 2 cm (f/# of 1) and a design wavelength of 4.5 μm. The metalens has a 10-fold reduction in lens thickness compared with a commercial Ge refractive lens. Importantly, the metalens is not designed to eliminate chromatic aberration across the 3–5 μm band. Following the analysis in Fig. 2, although the hyperboloid's broadband output is blurred, it preserves sufficient spatial information across the MWIR band to make computational restoration possible. Additional details, including the phase response of the meta-atoms is provided in the Supplementary Note 1.

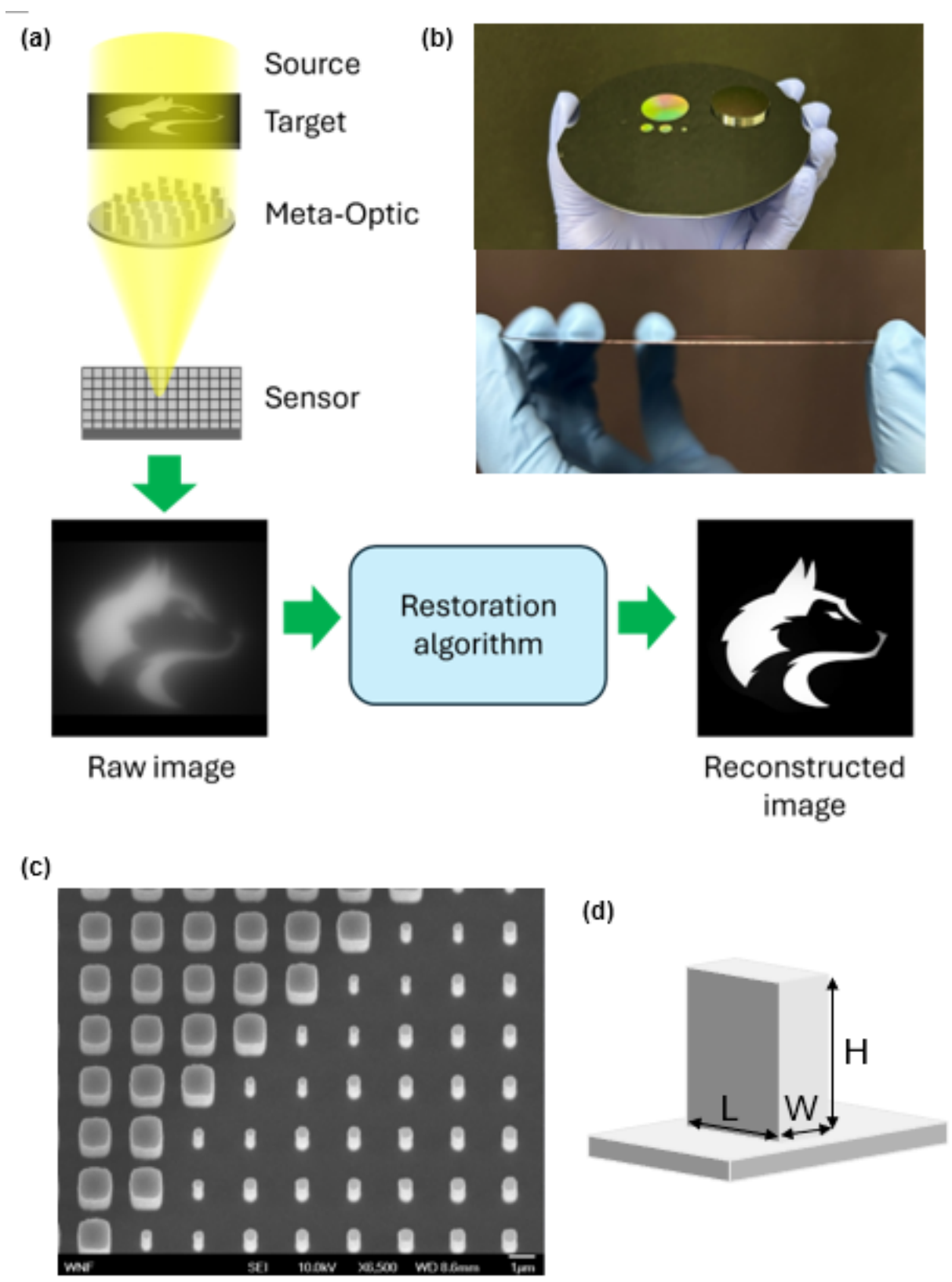

**Fig. 1. Overview of the proposed broadband MWIR imaging system with an all-Silicon hyperboloid metalens. (a)** Schematic description of the broadband imaging pipeline. A hot-plate heated between 100 – 200°C produces the MWIR signal and the images captured by a fabricated hyperboloid metalens is recorded by a monochromatic MWIR camera. The chromatic defocus present in the imaging measurements is restored using the computational restoration framework described in Supplementary Note 3. **(b)** Comparison of the bulk reduction from the Silicon-based metalens over a commercial off the shelf (COTS) Ge refractive lens. Side-perspective image of the ultra-thin form factor with all-Silicon fabricated metalens. **(c)** SEM of the fabricated all-Si meta-atoms. **(d)** Schematic of a meta-atom pillar. The height is 4 μm and the periodicity is 2 μm.

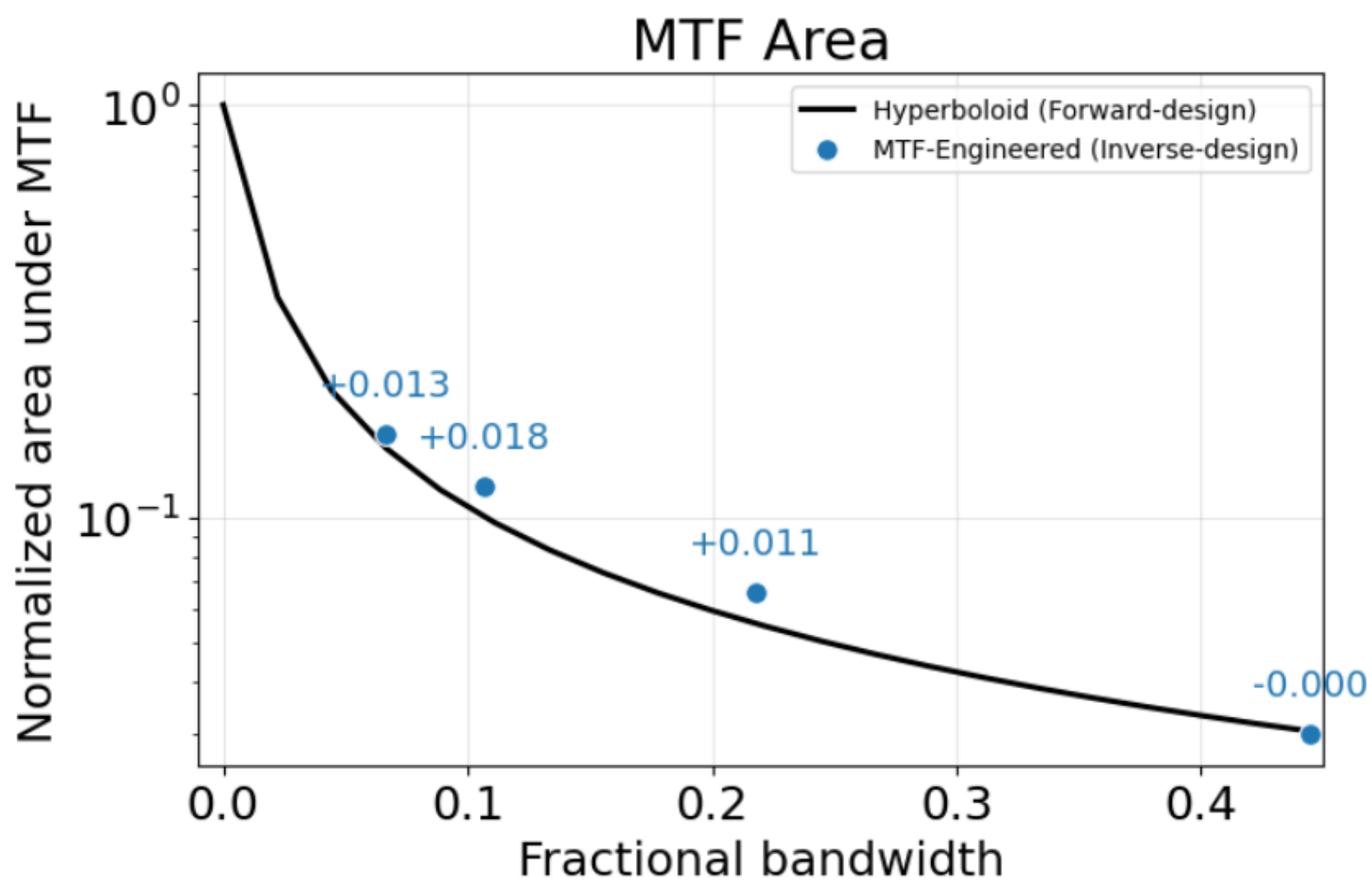

**Fig. 2. Comparison of simulated MTF performance of forward and inverse meta-optic designs.** Simulated integrated modulation transfer function (MTF) area as a function of fractional bandwidth for the matched hyperboloid (forward design) and MTF-engineered (inverse design) meta-optics. The integrated MTF area provides a comparison metric of broadband imaging performance across increasing fractional bandwidth. Blue annotations indicate the relative difference of the integrated MTF area between the forward- and inverse-designed meta-optics at matched optimization bandwidths. Across the full MWIR range (3 – 5 µm, fractional bandwidth = 0.5), the hyperboloid metalens exhibits less than a $10^3$ relative difference in MTF performance to the inverse design optic, suggesting that a forward design optic can achieve imaging performance comparable to an inverse design counterpart under identical aperture and focal length constraints.

**Few-shot computational restoration**

The simulation results in Fig. 2 establish that the hyperboloid metalens preserves sufficient broadband spatial contrast at the full MWIR range that could enable successful computational image restoration. However, this information is not directly accessible from the raw image because of strong chromatic blur. We therefore developed a computational restoration pipeline to determine whether the information retained by the metalens can be recovered experimentally (Fig. 3).

Our approach adapts an all-in-one image restoration model[33] to the MWIR imaging system. The model is pretrained on 16 common image-degradation tasks, including deblurring, dehazing, and low-light enhancement (Fig. 3f). This pretrained restoration prior is particularly useful for MWIR imaging, where fabrication imperfections, incoherent illumination, and sensor readout errors can cause experimental measurements to differ from numerical simulations. Rather than learning image restoration entirely from MWIR data, the model begins with previously learned representations of common image degradations and is then adapted to our experimental system.

Our image recovery pipeline consists of two stages: Stage 1 uses a few-shot data collection procedure to circumvent the lack of high-quality paired MWIR training data. Stage 2 finetunes the all-in-one restoration model and applies a refinement step to better match the MWIR target images. This approach allows the pretrained network to be adapted to our MWIR system using only four captured images.

We evaluate the adapted network on a separate, previously unseen, set of heated stencil targets (Fig. 4). Additional measurements using a 1951 USAF resolution test chart and exposure-bracketed captures are provided in the Supplementary Information. Across the broadband targets, the reconstructed images show substantially improved contrast and clearly recovered object structure. These results support the prediction from the optical analysis that the chromatically blurred measurements produced by the hyperboloid metalens still contain spatial information that can be recovered computationally, making it a competitive and practical approach to compact broadband MWIR imaging.

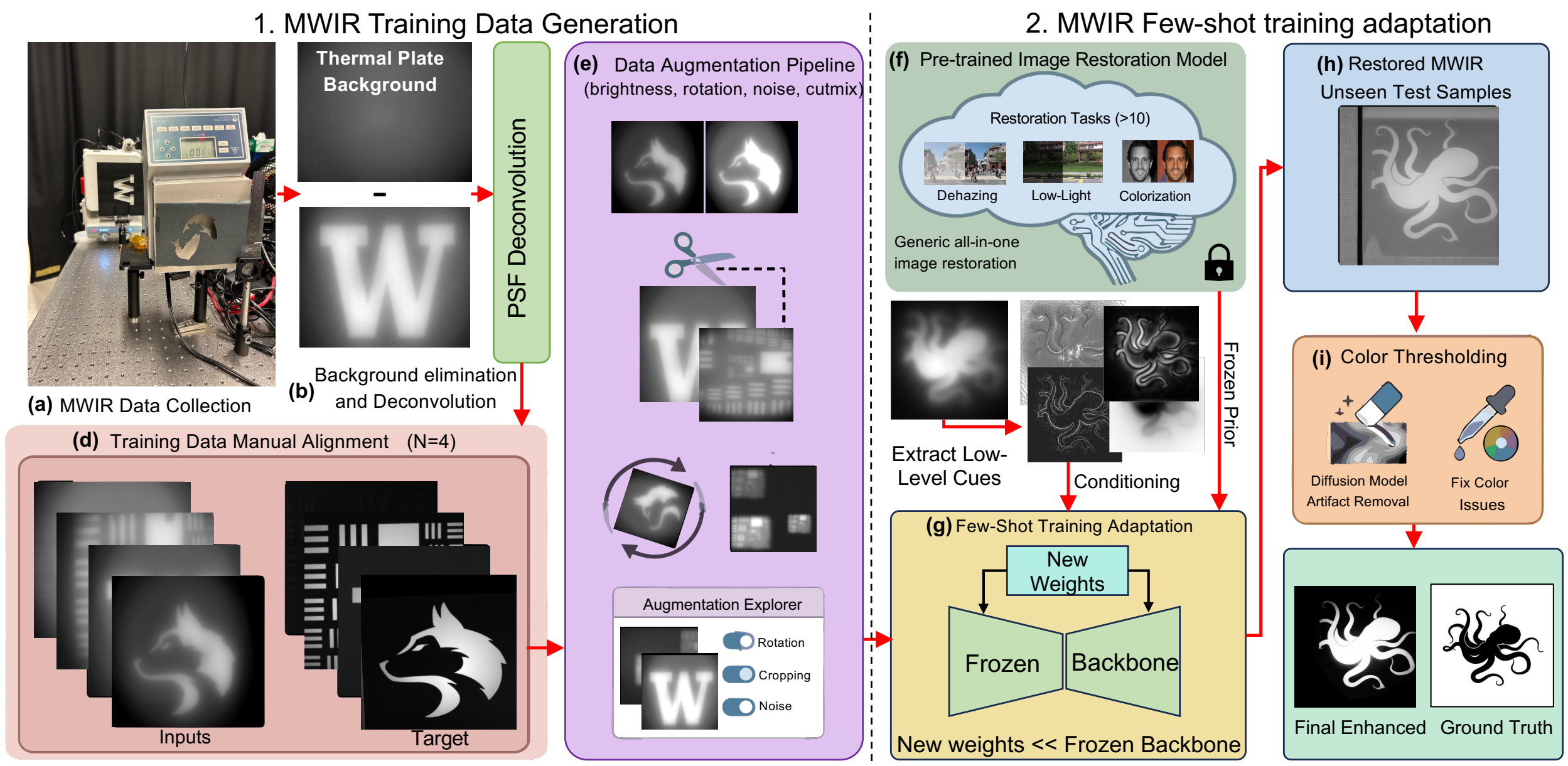

**Fig. 3. Restoration Model Pipeline. (a)** Overview of the proposed restoration framework involving a generative all-in-one restoration model followed by color refinement and artifact removal. **(b-e)** Data preparation of the all-in-one restoration for the MWIR task with 4 image inputs. The restoration model is paired with a robust data augmentation pipeline. **(f-i)** Fine tuning of the restoration in a parameter efficient manner. Enhanced images are passed through a color thresholding stage to correct for generative artifacts and correct for luminance.

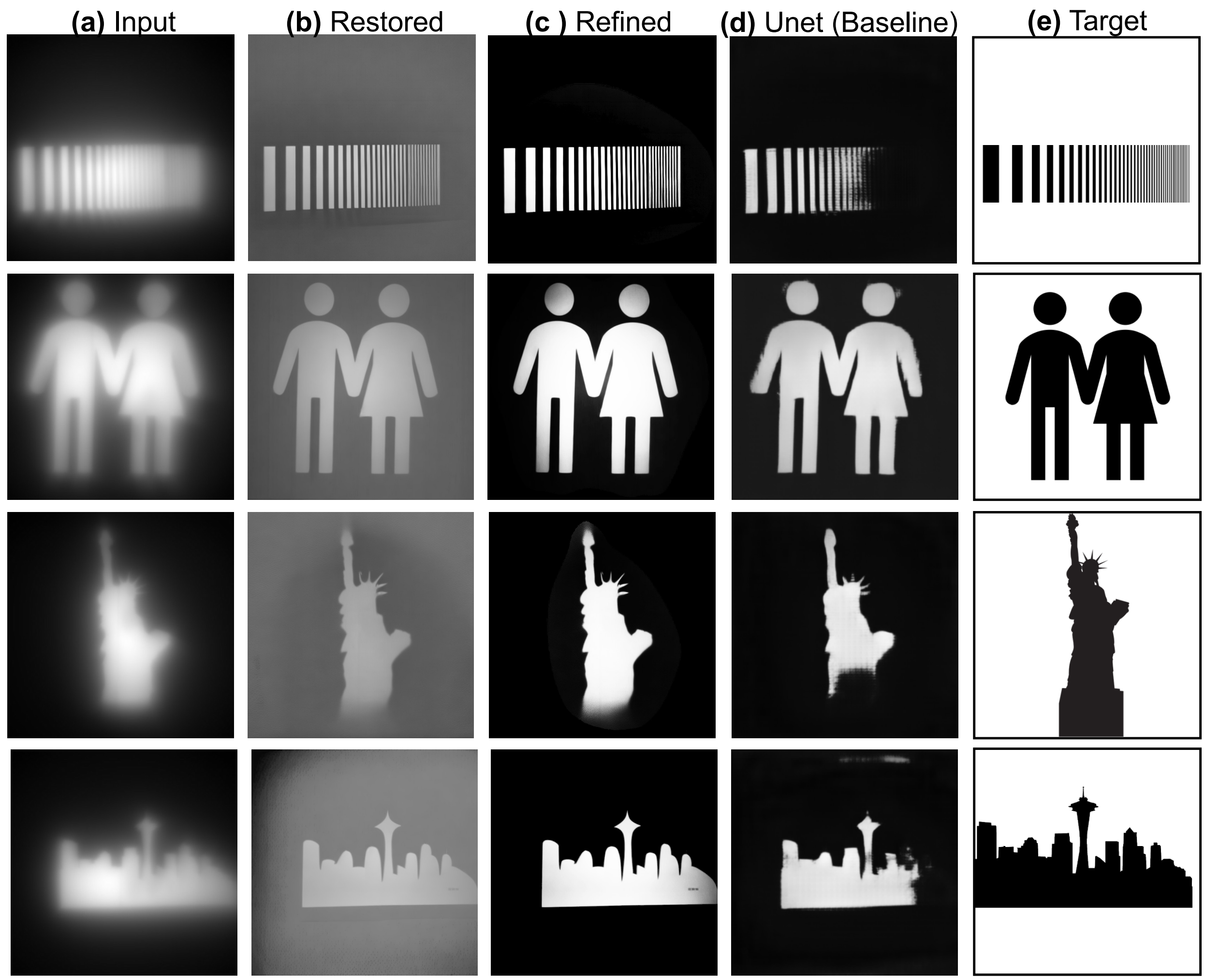

**Fig. 4. Imaging Results and Characterization. (a)** Broadband MWIR images with all-Silicon hyperboloid metalens using a monochromatic detector (FLIR A6751). **(b-c)** Restored images following the successive restoration and color refinement stages described in Methods. **(d)** Comparison against Unet-based baseline all-in-one restoration pipeline[34] which is unable to recover large scene regions. **(e)** Ground truth target images for comparison.

## Experimental broadband MWIR imaging

Figure 4a shows raw broadband images captured through the hyperboloid metalens, designed to focus at 4.5 μm wavelength, using a commercial FLIR camera. As expected, integrating light across the MWIR band produces strong chromatic blur. This blur reduces image contrast and suppresses high-spatial-frequency detail, making object boundaries difficult to distinguish directly in the raw measurements.

Figures 4b–d show the corresponding images after computational restoration. Despite the severe blur in the raw image captures, the restoration model successfully recovers broadband contrast from target objects with various identifiable structures and boundaries. The corresponding ground truth target images are shown in Figure 4e for comparison. The restored images show close agreement with the target structures, including the recovery of higher-spatial-frequency features associated with object edges and boundaries.

## Discussion

In this work, we systematically compare a single-wavelength hyperboloid metalens with an inverse-designed meta-optic across different fractional bandwidths at matched aperture and numerical aperture. Our findings demonstrate that broadband computational imaging does not necessarily require complex dispersion engineering or inverse-designed phase profiles. Across the full 3--5 $\mu$m MWIR band, a standard forward-designed hyperboloid metalens preserves sufficient spatial frequency information and contrast to enable high-fidelity computational image restoration.

This behavior is quantitatively governed by the relationship between fractional bandwidth and modulation transfer function (MTF) volume. Our analysis shows that the inverse-designed meta-optics achieves a larger integrated MTF volume within an intermediate fractional bandwidth regime ($\sim$ 0.10--0.20), matching the conditions in visible imaging where discrete color filters divide the spectrum into narrower sub-bands, i.e., RBG channels. Outside this regime, however, the hyperboloid retains higher overall MTF volume—exhibiting a 9% advantage across the full bandwidth ($\Delta\lambda = 2.0\,\mu$m) and a 3% advantage in the narrowband limit ($\Delta\lambda = 0.3\,\mu$m). For filter-less monochromatic MWIR detection, where a single channel collects light across a fractional bandwidth of 0.50, balancing the optical response over the entire spectral range no longer provides an advantage according to this metric. Instead, the simple hyperboloid preserves more of the spatial information available for computational reconstruction, making it a higher-performing optical front-end.

To recover this information experimentally, we adapt a pretrained image-restoration network to the measured response of our MWIR imaging system using only four captured images. This few-shot calibration avoids the large, paired datasets typically required to train image restoration networks from scratch. Despite the strong chromatic blur in the raw measurements, the reconstructed images show close agreement with the ground-truth targets. These experiments demonstrate that the information preserved by the hyperboloid metalens can be effectively recovered with limited system-specific training data.

More broadly, these results suggest that the optical front end need not eliminate chromatic aberrations if it preserves sufficient information for computational recovery across the broadband. This shifts the design objective away from producing an achromatic PSF and toward maximizing useful information transfer between the scene and sensor, thereby avoiding the fabrication complexity of multi-layer or multi-level metasurfaces. In our MWIR system, this principle enables broadband imaging using a simple, single-layer hyperboloid metalens without dispersion engineering, paired with a calibration-efficient computational backend. Such an approach provides a promising path toward compact, lightweight broadband MWIR imaging systems for SWaP-constrained sensing and imaging applications.

## Methods

Metalens fabrication

A 300 µm thick double-sided polished Silicon wafer was coated with a layer of HMDS to promote resist adhesion. A layer of positive resist (AZ 1505) was deposited (4k rpm, ~500 nm thickness), followed by a 60s soft bake at 100°C on a hot plate. The resist layer was then patterned with direct-write lithography (Heidelberg DWL66+), followed by a post-exposure bake for 60 s at 100°C on a hot plate. The resist was developed in a commercial developer solution (AZ 726) for 60 s. Subsequently, the resist pattern was transferred into the bulk Silicon with deep reactive-ion etching (SPTS DRIE) to a target depth of 3µm to form the Silicon nanopillars. The residual photoresist was stripped using oxygen plasma.

Measurements:

*PSF Measurement:* The refractive relay optics used to extend the imaging distance to the camera sensor plane were purchased from Edmund Optics. Specifically, a 50 mm diameter, 50 mm focal length Ge asphere (Edmund Optics #87-996) and 75 mm diameter, 150 mm focal length uncoated $CaF_2$ plano-convex lens (Edmund Optics, #24-821). To measure the PSF of the meta-optic, a tunable MWIR QCL is guided through a single mode fiber (Thorlabs IRFS32) to produce a point-source illumination ~25 cm away from the meta-optic. The PSF was then re-imaged with the refractive relay-optics and captured onto the camera sensor (FLIR MWIR A6751 InSb camera, 640 X 512 pixels, 15 µm pixel resolution).

*Images:* Broadband images were captured under blackbody radiation produced from a hot plate (Torrey Pines Scientific HP50) heated to 130°C. Laser-cut aluminum targets finished with matte black paint to prevent reflections were placed in front of the hot plate to pattern MWIR light and form the imaging object. Image captures were taken with the same refractive relay assembly and minimal post-processing routines, such as background subtraction, were performed before the reconstruction.

MWIR Meta-optic characterization

We fabricate a 20 mm aperture hyperboloid metalens with a nominal focal length of 20 mm with a central focusing wavelength of 4.5 μm. We compare this result against a known indicator for broadband imaging performance to an inverse designed meta-optic with matched dimensions of numerical aperture and focal length. The simulation results demonstrate that inverse design optics with intermediate optimization bandwidths outperform a hyperboloid, yet the forward design optics retain an advantage at the extremes of the bandwidth, specifically at 9% at the full $\Delta\lambda = 2.0$ μm band and 3% at the narrowest $\Delta\lambda = 0.3$ μm band. These are bandwidth ranges of interest for unfiltered thermal imaging and for narrowband gas sensing, respectively. Raw image measurements are captured with a commercial FLIR sensor (A6751 InSb) and a restoration algorithm enhances the monochromatic images; further details are in the Supplementary Material. A set of image captures from matte-black aluminum stencils are used to pattern contrast and the target intensity is recorded on the camera sensor without additional filters to emulate ambient, real-world conditions of randomly emitting hot objects.

Computational Reconstruction

(Stage 1) Data Collection: Due to practical limitations in collecting outdoors MWIR data, we build a robust data augmentation pipeline that can mimic outdoor conditions such as varying temperatures of scene objects, varying exposures, and varying object resolution scales based on distance from the sensor. Collecting paired data where the clean target is perfectly aligned to the corrupted MWIR input is also challenging due to the parallax between the regular and MWIR camera. We manually align lab-captured image pairs (Fig. 3.d) and use the best aligned 4-5 samples for the data augmentation pipeline (Fig. 3.e). The augmentation pipeline uses a library of augmentations ranging from simple geometric transformations such as rotation to complex image manipulations such as cutmix[35]. The resulting dataset is used for the all-in-one image restoration model finetuning phase described next.

(Stage 2) Diffusion Restoration Model: We build our restoration model upon a generative all-in-one image restoration framework coupling a frozen latent-diffusion backbone with task-specific mixture-of-experts control adapters[33]. Because the backbone and main control pathway remain frozen, the model can be specialized to a new degradation by training only a small set of task-isolated parameters, making it well suited to the data-scarce regime of MWIR metalens imaging. We first extract low-level cues such as edges, textures, haze maps, etc. from the input image as part of the base model's restoration pipeline. The restoration cue is used as conditioning additional to the source MWIR image. We introduce an additional lightweight branch (Fig. 3.g) for the MWIR restoration task which is finetuned on the augmented image dataset from Stage 1 with a small learning rate until performance saturates on a held-out frame, which restores the characteristic MWIR metalens aberrations (spatially varying blur and chromatic spread) while leaving the model's pre-trained restoration capabilities intact. This parameter-efficient scheme lets the generative prior of the base model carry most of the reconstruction burden, so that a very few samples (<5) suffice to recover a usable inverse for the metalens degradation.

Color Refinement Process: Our all-in-one diffusion restoration model was adapted to mid-wave infrared (MWIR) segmentation in a few-shot regime, fine-tuned on only four labelled samples. While the model generalized to unseen test samples at inference, outputs exhibit characteristic diffusion artifacts, spurious texture, halo regions around object boundaries, and intensity distributions that diverged from the target thermal scale (Fig. 3.h). A post-processing refinement pipeline (Fig. 3.i) was therefore developed to suppress these artifacts while preserving fine structural detail.

The color refinement pipeline applies three stages sequentially:

1. **Background suppression:** Pixels where the thermal input image pixels fall below a sensor floor ($\alpha$) were labelled as noise and zeroed, isolating true foreground thermal objects from sensor background.
2. **Sigmoid contrast remapping**: A sigmoid transform was centered on the per-image Otsu threshold[36] computed within the thermal foreground mask, pushing sub-threshold ($\beta_1$) pixels toward black and super-threshold ($\beta_2$) pixels toward white while preserving fine structural gradients.
3. **Edge-adaptive floor clipping**: Sobel edge detection[37] was then used to identify high-gradient regions in the sigmoid output. The $\gamma' th$ percentile of the strongest-edge pixels was used as a per-image floor, zeroing all values below it. This reduced spurious dark pixels with no loss of bright structure.

The hyper-parameters $\alpha, \beta, \gamma$ were tuned until the resulting distribution of restored images closely matched the target MWIR distribution.

**Data availability**

All relevant data that support the findings are available within this Article and Supplementary Information. Source data are available from the corresponding authors upon request.

**Code availability**

The diffusion model is available on:

**Acknowledgements**

This work was partially supported by NSF award 2107454, and AFOSR. Device fabrication and characterization work was conducted in part at the Washington Nanofabrication Facility / Molecular Analysis Facility, a National Nanotechnology Coordinated Infrastructure (NNCI) site at the University of Washington.

Distribution A: Approved for public release - distribution unlimited. (AFRL/RASE-2026-1791)

**Author contributions**

P.C. and A.M. conceived the idea. R.J. and Z.Z. designed the metasurfaces. D.M. and P.C. designed the simulation model and image restoration. R.J., A.W.S., and V.V. fabricated the devices. R.J. performed the experiments with the help from J.W. R.J., J.F., A.M., and P.C. interpreted the results together with all the authors. R.J. and D.M. wrote the manuscript with input from all the authors. A.M. and P.C. supervised the overall project.

**Competing interests**

The authors declare no competing interests.

# Supplementary Information for

# Broadband Mid-Wave Infrared Imaging Enabled by All-Silicon Hyperboloid Metalens

Rose Johnson[1,#] , Debrabrata Mandal[2,#], Zihao Zou[2], Jessie Wei[3], Anna Wirth-Singh[3], Vladimir Yarmolik[1], Johannes Froech[2], Arka Majumdar[1,3,*] Praneeth Chakravarthula[2,*]

[1] Department of Electrical and Computer Engineering, University of Washington, Seattle, WA 98195, USA.

[2] Department of Computer Science, University of North Carolina at Chapel Hill, Chapel Hill, NC 27517, USA.

[3] Department of Physics, University of Washington, Seattle, WA 98195, USA.

[*] Corresponding authors. Email: arka@uw.edu, cpk@cs.unc.edu

[#] These authors contributed equally

## Contents

<u>**Supplementary Notes**</u>

**Note 1. Details on the Hyperboloid Metalens Design**

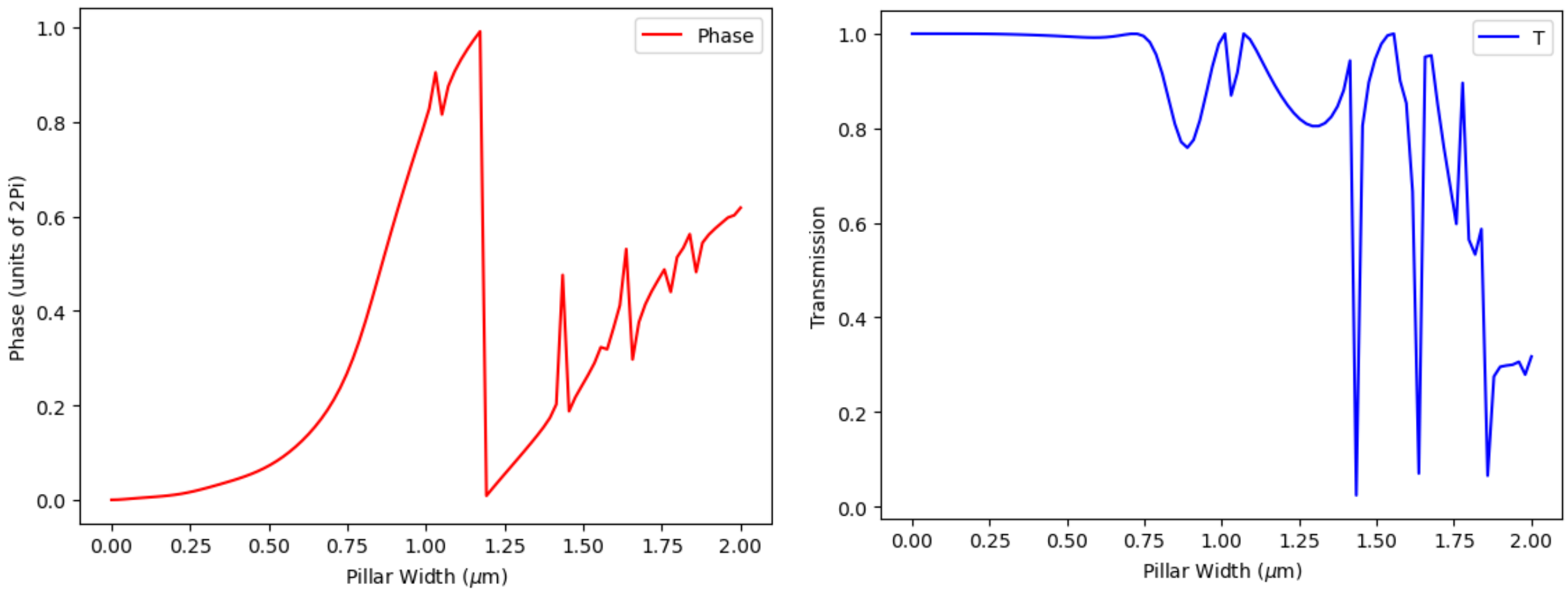

The scatterer (a) phase and (b) transmission responses are calculated using rigorous coupled wave analysis (RCWA) for pillar widths up to 2 μm. A discrete set of scatterers are selected with reasonable transmission to achieve the full 0 – 2π phase range. The dimensions are set to a height of 3 μm with a lattice periodicity of 2 μm. The wavelength is 4.5 μm.

The hyperboloid phase profile follows the equation,

$$\varphi(x, y) = -\frac{2\pi}{\lambda}(\sqrt{x^2 + y^2 + f^2} - f$$

Where $x, y$ are the position coordinates, $f$ is the lens focal length, and $\lambda$ is the nominal wavelength.

To fabricate the hyperboloid metalens, the calculated phase and transmission data is used as a lookup table to map the selected phase to corresponding pillar width. The fabricated metalens is all-Silicon with a 2 cm diameter and 2 cm focal length.

**Note 2. Integrated MTF Area Calculation**

The fractional bandwidth calculation includes an evaluation of imaging performance based on the simulated MTF area as function of spectral bandwidth. For a select number of bandwidths up to the full MWIR range, the wavelength range is centered around 4.5 μm and sampled every 10 nm. The fractional bandwidth considers the span of the wavelength range divided by the center frequency, and the monochromatic case is considered to be a bandwidth of 0.

To calculate the PSF, the angular spectrum method is used and the broadband PSF is obtained by integrating the non-normalized PSFs over the wavelengths across given bandwidth. The MTF is computed from the FFT of the PSF. Next, to find the MTF area, the MTF values are integrated between spatial frequency values between 0 to 100 cycles/mm across increasing bandwidths in the 10 nm intervals. Integration is performed using the numpy trapz function. The MTF area is normalized by the hyperboloid reference at zero bandwidth (monochromatic). This approach is completed for the hyperboloid phase profile and the four MTF-engineered phase profiles. A direct comparison is considered to find the difference between the MTF area between the MTF-engineered and hyperboloid at those specific bandwidth ranges.

**Note 3. Inverse Design Approach**

To perform the fractional bandwidth comparison, four inverse design profiles are simulated with the approach described in previous articles. The proxy model used a $8^{th}$ order harmonics for phase-to-structure mapping and $5^{th}$ (x = 5, y = 5) harmonic orders for the inverse structure-to-phase mapping. This process achieves a differentiable mapping between structure (duty cycle) and phase to fit the phase data of the metasurfaces to a polynomial proxy function,

$$d(r) = \sum_{i=0}^{N} b_i \left(\frac{\phi(r)}{2\pi}\right)^{2i}$$

Where $d(r)$ is the required duty cycle at a given position $r$ from the optical axis of the metasurface, $\phi(r)$ is the phase at the given position for the nominal wavelength $\lambda = 4.0\ \mu m$, and $b_i$ are fittting parameters.

The inverse design metalenses include an all-Silicon design which considers a 2 μm periodicity and 3 μm height. The diameter is 1.2 cm and the focal length is 2.5 cm across the full 3 – 5 μm wavelength range.

The proxy function is modeled by fitting the pre-computed transmission coefficients with a radially symmetric second-order polynomial function,

$$\tilde{\phi}(r,\lambda) = \sum_{n=0}^{2}\sum_{m=0}^{2} b_{nm} d(r)^n \lambda^m \,, n + m \leq 2$$

Where $\lambda$ is a non-nominal wavelength and $b_{nm}$ are fitting parameters. The transmission values are calculated using RCWA and fitted to the polynomial using linear least squares.

We initialize the radial phase profile $\phi(r)$ with a quadratic focusing phase and optimize its values at discrete radial positions. At each iteration, the proxy functions map $\phi(r)$ to the duty-cycle distribution $d(r)$ and predict the phase response $\tilde{\phi}(r,\lambda)$. For wavelengths sampled uniformly across the optimization band, the transmitted fields are propagated to the focal plane using the angular spectrum method implemented with a zeroth-order Hankel transform. The resulting PSFs are used to calculate the normalized MTFs.

The phase profile is optimized by minimizing the loss function,

$$\mathcal{L}[\phi] = -\frac{1}{N_\lambda N_\nu}\sum_{j=1}^{N_\lambda}\sum_{k=1}^{N_\nu} \mathrm{MTF}\left(\nu_k, \lambda_j; \phi\right),$$

where $N_\lambda$ is the number of sampled wavelengths and $N_\nu$ is the number of spatial-frequency samples below the optimization cutoff $\nu_c$. Gradients with respect to the sampled phase values are computed by automatic differentiation through the proxy functions and propagation model. The Adam optimizer updates the phase profile, and the procedure is repeated independently for each of the four optimization bands.

**Note 4. Computational Restoration Algorithm**

We implement a probabilistic image restoration framework for MWIR images, based on a generative all-in-one restoration model. Given a degraded MWIR image, the model first extracts low-level visual cues, such as edges, textures, and intensity information. The model has been pretrained on a large-scale dataset spanning multiple image degradation and restoration tasks, allowing it to recognize these common image features across different degradation types. This prior knowledge is particularly useful for MWIR restoration, where the captured images can differ substantially from the data encountered during pretraining. Although the MWIR images themselves may be out of distribution, their low-level features share common structure with those learned from other restoration tasks. Representing the input through these features therefore provides a more generalizable prior for restoring previously unseen MWIR degradations.

The extracted low-level cues are then used as conditioning information for a denoising diffusion-based restoration model. Within its latent space, the model identifies the degradations present in the input and combines learned degradation-specific expert adaptors to guide the restoration process. This allows the model to draw on restoration strategies learned from multiple degradation types rather than relying on a model trained specifically for the MWIR imaging system. Finally, the restored output undergoes a refinement stage to reduce artifacts introduced by the domain gap between the pretraining and MWIR datasets. Together, these stages allow a pretrained restoration model to generalize to the chromatic degradation produced by our MWIR metalens while requiring only limited system-specific adaptation.

## Supplementary Figures

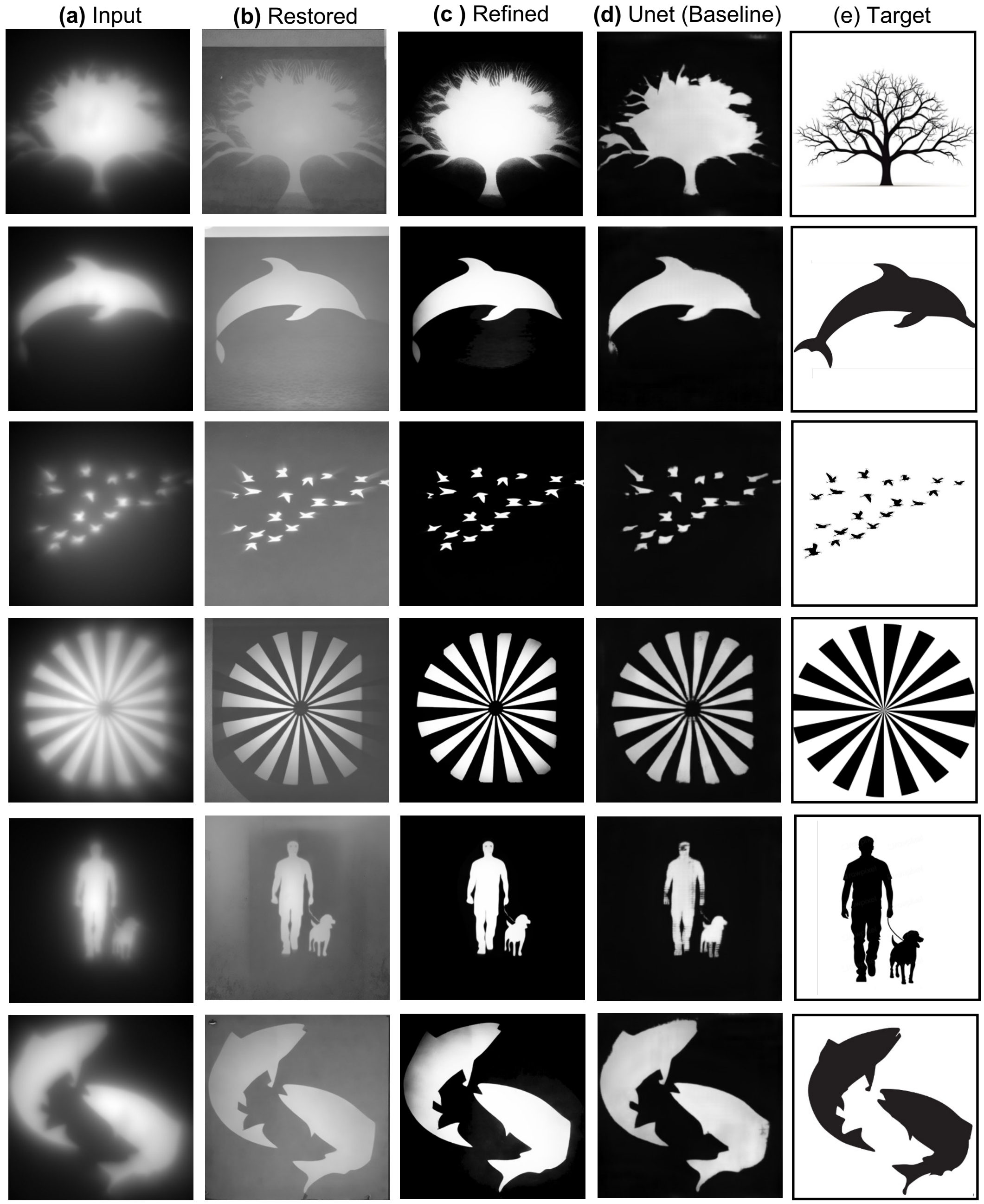

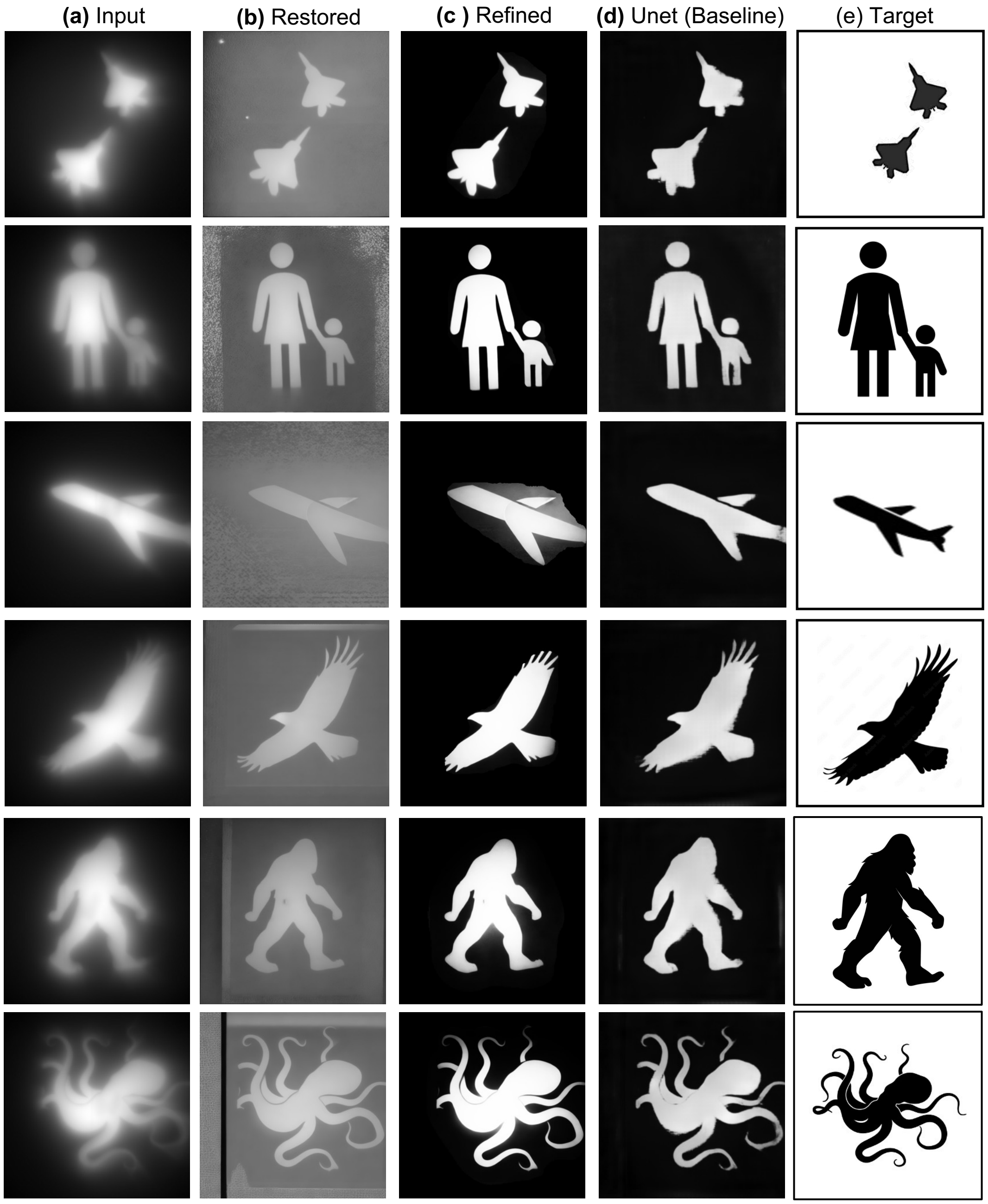

**Fig. S1. Additional MWIR broadband imaging results from varying targets.**

Experimental imaging results from additional target stencils captured by the hyperboloid metalens (input) with the restored image results compared to the ground truth baseline. **(a)** Captured images using a monochromatic detector (FLIR A6751). **(b-c)** Restored images following the successive restoration and color refinement stages described in Methods. **(d)**

Comparison against Unet-based baseline restoration pipeline. **(e)** Ground truth target images for comparison.

While there are some artifacts that appear as a result from the computational imaging reconstruction pipeline, there is broadly an increase in the broadband intensity contrast produced between the input and refined image sets with additional restoration of fine feature details particularly at the target stencil boundaries.

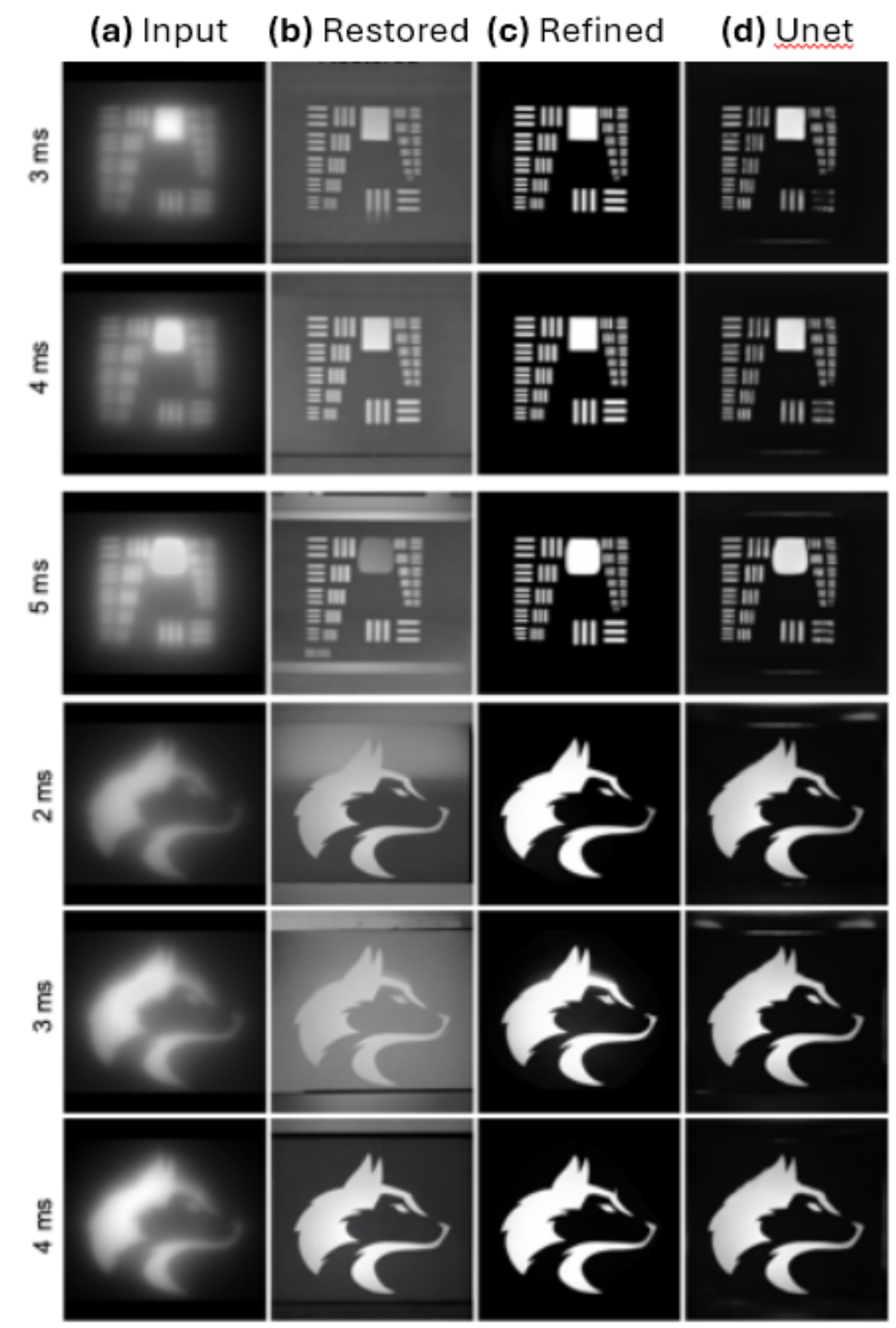

**Fig. S2. MWIR broadband imaging results from exposure bracketed captures**

Experimental imaging results from additional target stencils captured by the hyperboloid metalens (input) with the restored image results compared to the ground truth baseline at increasing exposure times. **(a)** Captured images using a monochromatic detector (FLIR A6751). **(b-c)** Restored images following the successive restoration and color refinement stages described in Methods. **(d)** Comparison against Unet-based baseline restoration pipeline**.**

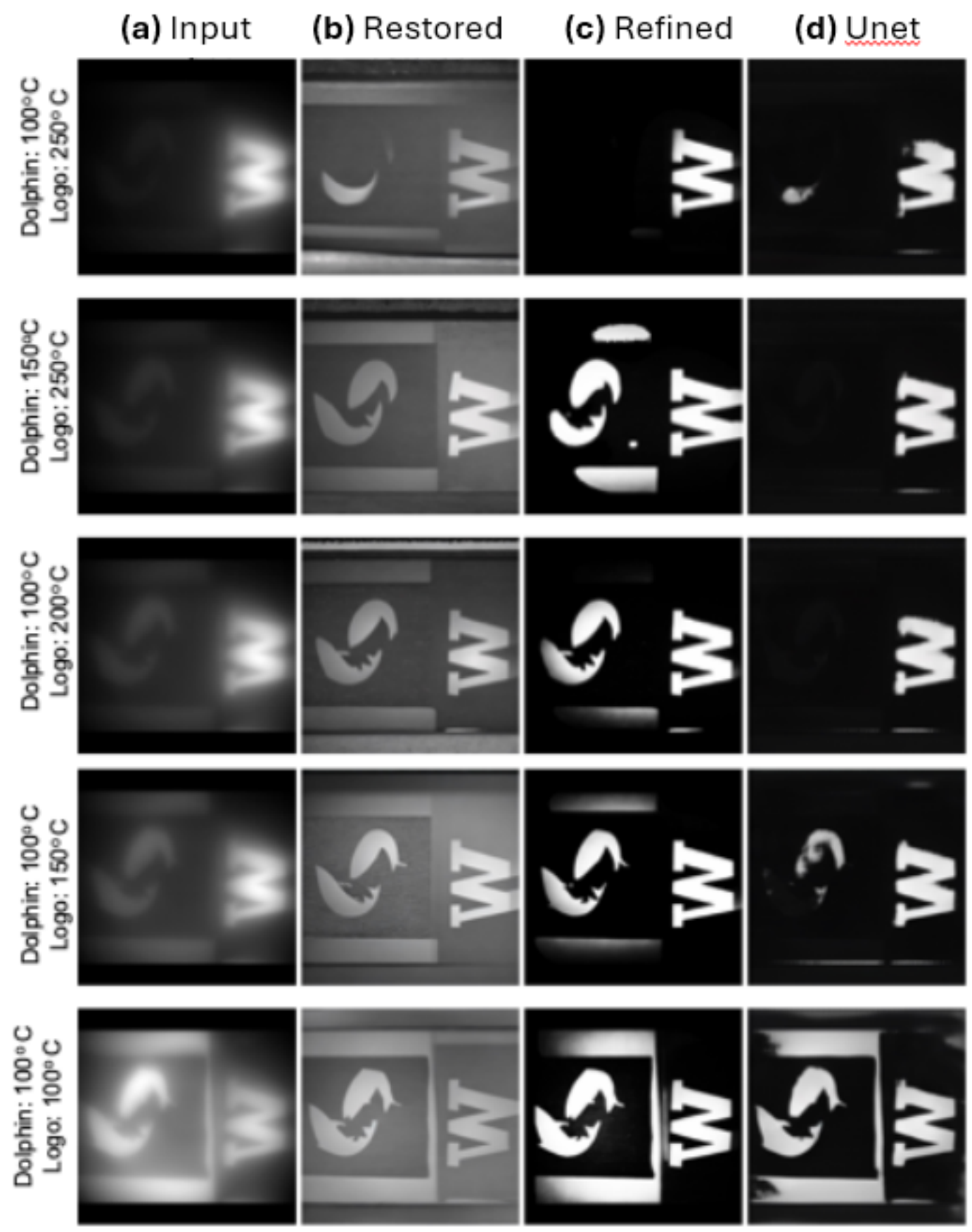

**Fig. S3. MWIR broadband imaging results under varying temperatures**

Experimental imaging results from target stencils captured by the hyperboloid metalens (input) with the restored image results compared to the ground truth baseline at different stencil temperatures. Two stencils are imaged at different temperatures ranging from 100-250$^{o}$ C and the resulting image is captured in the MWIR camera. The output is restored using both our method and compared against the Unet based baseline method.